\documentclass{article}
\usepackage{amsmath}
\usepackage{amssymb}
\usepackage{bm}
\usepackage{graphicx}
\usepackage{color}
\begin{document}
\begin{center}
\LARGE{Two dimensional representation of the Dirac equation in
Non associative  algebra }
\end{center}
\medskip
\begin{center}
S. Hamieh$^{1}$, and H. Abbas$^{2}$

$^{1}$Department of Physics, Lebanese University Faculty of Science,
Beirut Lebanon\\
$^{2}$Department of Mathematics Lebanese University Faculty of Science, Beirut  Lebanon
\end{center}
\begin{quote}

In this note a simple extension of the complex algebra to  higher
dimension is proposed. Using the proposed algebra a two
dimensional  Dirac equation is formulated and its solution is calculated.  It is found that
there is a sub-algebra where the associative nature can be recovered.

\end{quote}

\section{Introduction}
 The physical motivation
of a generalized quantum mechanics is that, although the
low-energy effective theories governing the strong, electroweak,
and gravitational interactions of elementary particles are
believed to be described by local complex quantum field theories,
attempts to construct an underlying unifying theory within the
same framework have run into difficulties. Perhaps a successful
unification of the fundamental forces will require one or more new
ingredients at the conceptual level. One possibility,
 is to sacrifice the assumption of locality or of
"point" particles, as is done in string theories. A second
possibility, which motivates the present work, is that a
successful unification of the fundamental forces will require a
generalization beyond complex quantum mechanics\cite{Adler,1936,Finkelstein2,Yefremov,Frenkel,Negi,Leo1,Khalek,Adler1,Maia,Rawat,Rawat2}.

 In the present
work, our aim, is to
give a description of an algebra which can be used in a possible extension of the  local complex quantum field theories.
Also, a considerable emphasis is placed on the development of
two dimensional Dirac equation.
A number of interesting and characteristic features of
 the non associative algebra will be seen to emerge.

This paper is organized as follow: in the next  section the description of the algebra is given. Section 3 is devoted to the study of two dimensional Dirac equation.  Our conclusion will be given  in section 4.

\section{Number systems used in Quantum mechanics and the Generalized-$\mathbb{C}$}

 To determine the allowed structure of the algebra that can be used for
a generalized quantum mechanics, Adler \cite{Adler}
 introduced a number of assumptions concerning the properties of the
modulus function $N$ of the number $\phi$ elements
of the  algebra: { \begin{equation}   N(0)=0\label{eq1}\end{equation}
\begin{equation}N(\phi)>0\quad \rm if\, \phi\neq 0\label{eq2}\end{equation}
\begin{equation}N(r\phi)=|r|N(\phi)\quad r\,\rm real\label{eq3}\end{equation}
\begin{equation}N(\phi_1+\phi_2)\leq N(\phi_1)+N(\phi_2)\label{eq4}\end{equation}
\begin{equation}N(\phi_1\phi_2)= N(\phi_1)N(\phi_2)\label{eq5}\end{equation}

The $\phi$'s  are
elements of a general finite dimensional algebra over the real
numbers with unit element, of the form
\[\phi=\sum_A r_A e_A\]

where $r_A$ are real numbers and the $e_A$ are basis elements
of the algebra, obeying the multiplication law
\[e_Ae_B=\sum_C f_{ABC}e_C\]
with real-number structure constants $f_{ABC}$.
}

 With the help of Albert theorem \cite{Albert}, it is found that
the only algebras over the reals, admitting a
modulus function with Adler properties, are
the reals $ \mathbb{R}$, the complex numbers $ \mathbb{C}$, the
quaternions or Hamilton numbers $ \mathbb{H}$ and the octonions or
Cayley numbers $ \mathbb{O}$.

However, from experimental point of view, there is no guarantee that the Adler postulate about
the modulus function $N$ will  be satisfied in the new energies domains. Perhaps, new physics can emerge
\cite{Hamieh}.
Thus, there should be no restriction about the algebra that can be used for a possible
extension of the complex algebra. The only requirement is that the expected extension 
 should verify  Adler postulate  in its
sub-algebra. Moreover,
it is more natural, to assume a simple extension,  rather than making extension to 4 dimensional or even  higher.
This idea will be used in our approach. In fact,  we propose to use a three dimensional algebra.
 Our intuitive assumption is
based on a geometrical approach as proposed by Descartes in
describing the complex number.

\underline{\bf Our Approach:} The proposed generalization  of the $\mathbb{C}$ algebra, the
Generalized-$\mathbb{C}$ (G$\mathbb{C}$), is finite-dimensional
non division algebra\footnote{ A division algebra,  is a finite
dimensional algebra for which $a\neq0$ and $b\neq 0$ implies
$ab\neq  0$, in other words, which has no nonzero divisors of
zero. }
 containing the real numbers $\mathbb{R}$ as a
sub-algebra and  has the following properties:
\begin{itemize}
\item A general G$\mathbb{C}$ number, \emph{q}, can be written as
\[ q = a + bi + cj \]
where
\[\emph{a,b,c,}
\in \mathbb{R} \,\rm or\, \mathbb{C}\,,\]
and the imaginary G$\mathbb{C}$ units, $i, j$ are defined by
\[ii = jj = -1\]
\[ij = ji = 0\]

\item The
addition  defined as
\[q_1 + q_2 = a_1 + a_2 +(b_1 + b_2)i + (c_1 +c_2)j\,, \] is associative
\[ q_1 + (q_2 +q_3) = (q_1 + q_2) + q_3\,.\]
\item The multiplication defined  as
\[q_1q_2 = (a_1a_2 - b_1b_2 - c_1c_2 ) \]
\[+(a_1b_2 + b_1a_2)i \]
\[+(a_1c_2 + c_1a_2 )j \,,\]

is non-associative under
multiplication that is $(q_1q_2)q_3 \neq q_1(q_2q_3)$.
\item The norm of an element $q$ of  G$\mathbb{C}$ is defined by
\[N(q)=(\bar{q}q)^{1/2}=(a^2 + b^2 + c^2)^{1/2}\]
with the   G$\mathbb{C}$ conjugate $q$ given by
\[\bar{q}=a - bi -cj\,.\]
\end{itemize}
Using the properties of the G$\mathbb{C}$,  a generalization of the Euler formula to three dimension can
be found. In fact any $q$ can be written as
\[ q = a + bi + cj =Re^{\theta(\cos(\phi)i+\sin(\phi)j)}=R\sum{(\theta(\cos(\phi)i+\sin(\phi)j))^n\over n!}\]
where
$a=R\cos(\theta),\,b=R\sin(\theta)\cos(\phi),\,c=R\sin(\theta)\sin(\phi)$
and ${R,\theta,\phi}
\in \mathbb{R}$, are the distance from the origin, the polar and the azimuthal
angle in the three dimensional Euler space.\\

{\bf Note that}, $e^{\theta(\cos(\phi)i+\sin(\phi)j)}\neq e^{\theta\cos(\phi)i}
e^{\theta\sin(\phi)j}$ because
the multiplication law is commutative but not assciative.

It is important to note, as we mention previously, that
there exist a sub-algebra of G$\mathbb{C}$ where the probability is preserved
in quantum mechanics. In this sub-algebra we must
assume that the azimuthal phase $\phi$ is constant. In this case
G$\mathbb{C}$ will be an associative and division sub-algebra that we
call  special G$\mathbb{C}$ (SG$\mathbb{C}$). Thus  any two numbers in this sub-algebra can be
written as
\[ q_1 = a_1 + b_1i + c_1j =R_1e^{\theta_1(\cos(\phi)i+\sin(\phi)j)}\,,\quad q_2 = a_2 + b_2i + c_2j =R_2e^{\theta_2(\cos(\phi)i+\sin(\phi)j)}\]
where the phase $\phi$    is a free parameter that can be
determined from physical properties. Also in this sub-algebra
the product of two elements have a physical meaning that is
a rotation in the Euler space
\[q_1q_2=R_1R_2e^{(\theta_1+\theta_2)(\cos(\phi)i+\sin(\phi)j)}\,.\]

\section{ Two dimensional Dirac's Equation in the Generalized-$\mathbb{C}$
 } 
 In this section
an \emph{ab initio} development of the Dirac formalism in two
dimension using the proposed
Generalized-$\mathbb{C}$ is discussed. In fact, in $\mathbb{C}$, Dirac's equation is
often given as
\[(i\gamma^\mu\partial_\mu - m)\psi = 0\]
which involves $i \in \mathbb{C}$ and thus forces the first
decision point in transitioning to another mathematical algebra.  For
clarity, to avoid the explicit use of $i$, the most general form
($c = \hbar = 1$) of Dirac's equation is
\begin{equation}
\mathcal{H}\psi=(C_\mu\partial_\mu)\psi=(C_x \partial_x + C_y \partial_y + C_z \partial_z + C_t \partial_t)\psi= m\psi. \label{eqn:dw}
\end{equation}
To recover the Klein-Gordon equation
\begin{equation}
(\nabla^2 - \partial_t^2)\psi =  m^2\psi , \label{eqn:KG}
\end{equation}
the following conditions must hold
\begin{equation}C_{x,y,z}^2 = 1;\hspace{.1in}  C_t^2 =  -1; \hspace{.1in}\mbox{ and }\left\{C_\mu,C_\nu\right\}=C_\mu C_\nu + C_\nu C_\mu =0, \mbox{ where } \mu\neq\nu.\hspace{.1in}\mu,\nu = x,y,z,t\label{eqn:cnd}
\end{equation}
 \\
Equation ($\ref{eqn:dw}$) can be rewritten as\footnote{Throughout this work the position of the indices, $\mu, \nu$ etc have no significance with respect to covariance or contravariance and are placed for typographical convenience.  Repeated indices, however, do indicate summation}
\begin{equation}
(\gamma^\mu\partial_\mu -m)\psi = 0,\hspace{.3in}\mu=0,1,2,3 \label{eqn:qb}
\end{equation}
by defining
\[\gamma^\mu = (C_t,C_x, C_y, C_z)\]  this avoids the explicit use of an imaginary scalar.
Using the following Dirac matrices, that take into account  $\emph{\;i,\;j\;}$ symmetry,
satisfying ($\ref{eqn:cnd}$)  \vspace*{.1in}

{\centering
$\gamma^0\equiv \gamma^t \equiv C_t = \left(
\begin{array}{cccc}
0 & j \\
j & 0
\end{array} \right)
$ \vspace*{.1in}

$\gamma^1 \equiv \gamma^x \equiv C_x = \left( \begin{array}{cccc}
 0 & -j \\
 j & 0
      \end{array} \right)$
\vspace*{.1in}

$\gamma^2 \equiv \gamma^y \equiv C_y = \left( \begin{array}{cccc}
0 & -i \\
i & 0
      \end{array} \right)
$ \vspace*{.3in}

$\gamma^3\equiv \gamma^z \equiv C_z = \left(
\begin{array}{cccccccc}
 1& 0 \\
0 & -1
       \end{array} \right)
$\\

}

 in
equation($\ref{eqn:qb}$) results in: \vspace*{.2in}
\begin{equation}
(\mathcal{H}-m) \psi=
\left(\begin{array}{cccc}
-m+\partial_z & -j\partial_x -i \partial_y + j\partial_t \\
j\partial_x + i\partial_y +j\partial_t & -m-\partial_z
\end{array}
\right)\left(\begin{array}{c} \psi_1 \\ \psi_2
\end{array}\right)=0 \label{eqn:dm}
\end{equation}
\vspace*{.13in}
\newline
The solution to this equation in 1+1 dimension, $x,t$, can be found
\begin{equation}
\psi(x,t)= N\left(\begin{array}{c} {E+p\over m} \\1
\end{array}\right)e^{j(px-Et)}\quad \,  \label{eqn:dm}
\end{equation}

as usual  $p$ represent the `momentum' and $E=\pm \sqrt{p^2+m^2}$ is the `energy' and
$N$ is a normalization factor.

\begin{figure}
\centering\includegraphics[width=12cm]{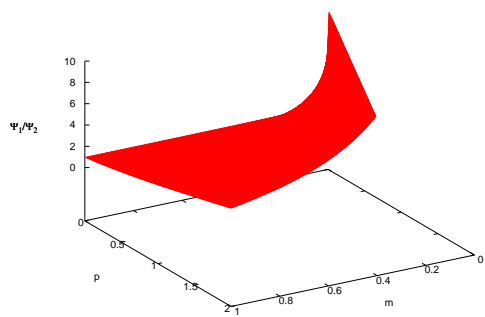}
\vspace{0cm}\caption{$\Psi_1/\Psi_2$ as function of $m$ and $p$.
\label{conc}}
\end{figure}

The ratio $\psi_1/\psi_2$ is shown in
Fig. 1. Discussion about 
the physical meaning of these results  and application to local field theory will be left for future investigation. 

\section{Conclusions}

We have found a two dimensional Dirac wave function in non a
associative algebra. A possible criticism of the approach of this
paper is that is there a physical quantities that can emerge? The
intent of future articles is to contribute to the resolution of
that debate. Finally  we believe that such algebra
merit to be explored in more
physical problems.\\

\end{document}